\documentclass[journal]{IEEEtran}

\usepackage{cite}
\usepackage{amsmath}
\usepackage{graphicx}

\newcommand{\ud}{\mathrm{d}}
\newcommand{\eqn}[1]{(\ref{#1})}
\newcommand{\fig}[1]{Fig. \ref{#1}}
\newcommand{\Fig}[1]{Figure \ref{#1}}
\newcommand{\Figs}[1]{Figures \ref{#1}}
\newcommand{\figs}[1]{Figs. \ref{#1}}

\begin{document}

\title{Error Probability of DPSK Signals with Intrachannel Four-Wave-Mixing in Highly Dispersive Transmission Systems}

\author{Keang-Po Ho,~\IEEEmembership{Senior Member,~IEEE}%
\thanks{Manuscript received October ??, 2004, revised ??, 2004. %
This research was supported in part by the National Science Council of Taiwan under Grant NSC-93-2213-E-002-061 and NSC92-2218-E-002-055}
\thanks{K.-P. Ho is with the Institute of Communication Engineering and Department of Electrical Engineering, National Taiwan University, Taipei 106, Taiwan.
(Tel: +886-2-2363-6262 ext. 222, Fax: +886-2-2368-3824, E-mail: kpho@cc.ee.ntu.edu.tw)}
}

\markboth{IEEE Photonics Technology Letters}{K.-P. Ho: Error Probability of DPSK Signals with Intrachannel Four-Wave-Mixing in Highly Dispersive Transmission Systems}

\maketitle

\begin{abstract}
A semi-analytical method evaluates the error probability of DPSK signals with intrachannel four-wave-mixing (IFWM) in a highly dispersive fiber link with strong pulse overlap. 
Depending on initial pulse width, the mean nonlinear phase shift of the system can be from 1 to 2 rad for signal-to-noise ratio (SNR) penalty less than 1 dB.
An approximated empirical formula, valid for penalty less than 2 dB, uses the variance of the differential phase of the ghost pulses to estimate the penalty.
\end{abstract}


\section{Introduction}

\PARstart{D}{ifferential} phase-shift keying (DPSK) signal has been studied widely recently for long-haul lightwave transmission systems \cite{gnauck02, zhu03, rasmussen03, cai04}.
In additional to the 3-dB receiver sensitivity improvement to on-off keying (OOK), high-speed DPSK signal also has larger tolerance to fiber nonlinearities than OOK signal \cite{xu04}.
Most DPSK experiments use return-to-zero (RZ) short pulse and launch a constant-intensity pulse train with phase modulated to each RZ pulse.

For 40-Gb/s signal in dispersive fiber, each RZ pulse broadens very fast by chromatic dispersion and overlaps with each other.
The pulse-to-pulse interaction gives intrachannel cross-phase modulation (IXPM) and four-wave-mixing (IFWM) \cite{shake98, essiambre99}.
As a constant pulse train, IXPM induces identical phase modulation and timing jitter to all pulses and does not affect DPSK signal.
However, IFWM adds ghost pulses to each DPSK RZ pulse \cite{mecozzi00, mecozzi01a,abo2, wei03fwm}. 
DPSK signal also has higher tolerance to IFWM than OOK signal \cite{wei03fwm}.

When the IFWM induced ghost pulses are evaluated numerically, the error probability of DPSK signal can be calculated semi-analytically.
This letter studies the statistical properties of IFWM in more detail using a method similar to \cite{mecozzi01a, wei03fwm}.
Both the error probability and signal-to-noise ratio (SNR) penalty are calculated.

\section{Statistics of IFWM}

If the signal launched to the fiber link is Gaussian pulse train with initial $1/e$-pulse width of $T_0$ or full-wide-half-maximum (FWHM) pulse with of $1.66 T_0$, for a bit-interval of $T$, the $k$th pulse is $u_k = A_k \exp\left[-(t - kT)^2/2T_0^2 \right]$, where $A_k = \pm A_0$ is phase modulated by either $0$ or $\pi$.
From \cite{mecozzi00, mecozzi01a, wei03fwm}, the peak amplitude of the ghost pulses induced from IFWM, from the $m$, $n$, and $(m+n)$th pulses to the pulse at $t = 0$ is 
\begin{eqnarray}
& &\Delta u_{m, n} =  i \gamma A_m A_n A^*_{m+n} \nonumber \\
& & \quad \times \int_{0}^L \frac{e^{-\alpha z}}{\sqrt{1 + 2 j \beta_2 z/T_0^2 + 3(\beta_2 z/T_0^2)^2}}  \nonumber \\
& & \quad \quad \times \exp\Bigg\{
    - \frac{3 m n T^2}{T_0^2 + 3j \beta_2 z} \nonumber \\
& & \quad \quad \quad - \frac{(m-n)^2 T^2}{
           T_0^2\left[ 1 + 2 j \beta_2 z/T_0^2 + 3(\beta_2 z/T_0^2)^2\right]}
 \Bigg\} \ud z,
\label{deluifwm}
\end{eqnarray}

\noindent where $m \neq 0$ and $n \neq 0$, $\gamma$ is the nonlinear coefficient and $\alpha$ is the attenuation coefficient of the fiber, $\beta_2$ is the coefficient of group velocity dispersion, and $L$ is the fiber length per span.
Here, we exclude both IXPM with either $m = 0$ or $n = 0$ and self-phase modulation with $m = n = 0$.

\begin{figure}
\begin{center}
\begin{tabular}{cc}
\includegraphics[width = 0.22 \textwidth]{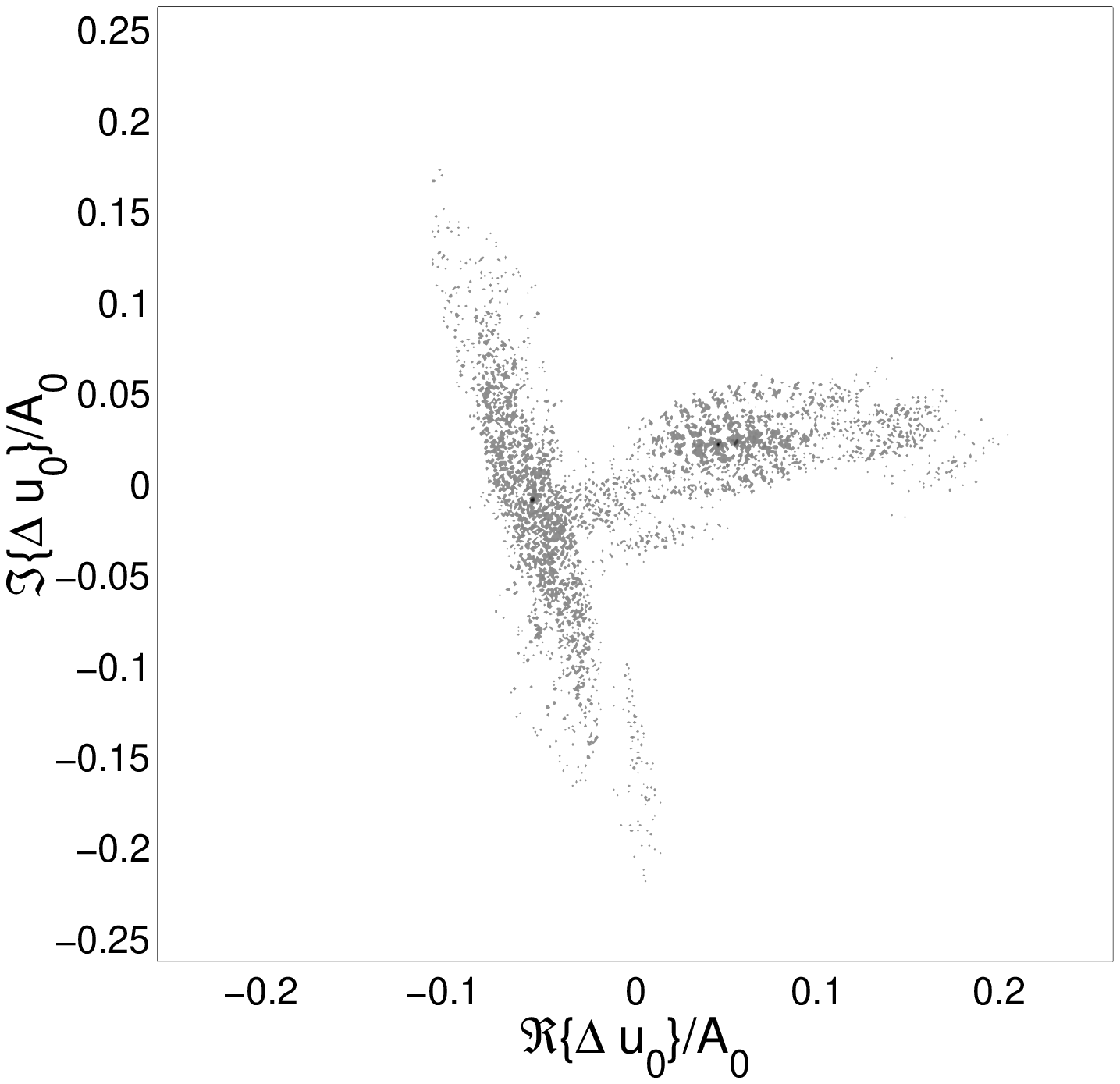} &
\includegraphics[width = 0.22 \textwidth]{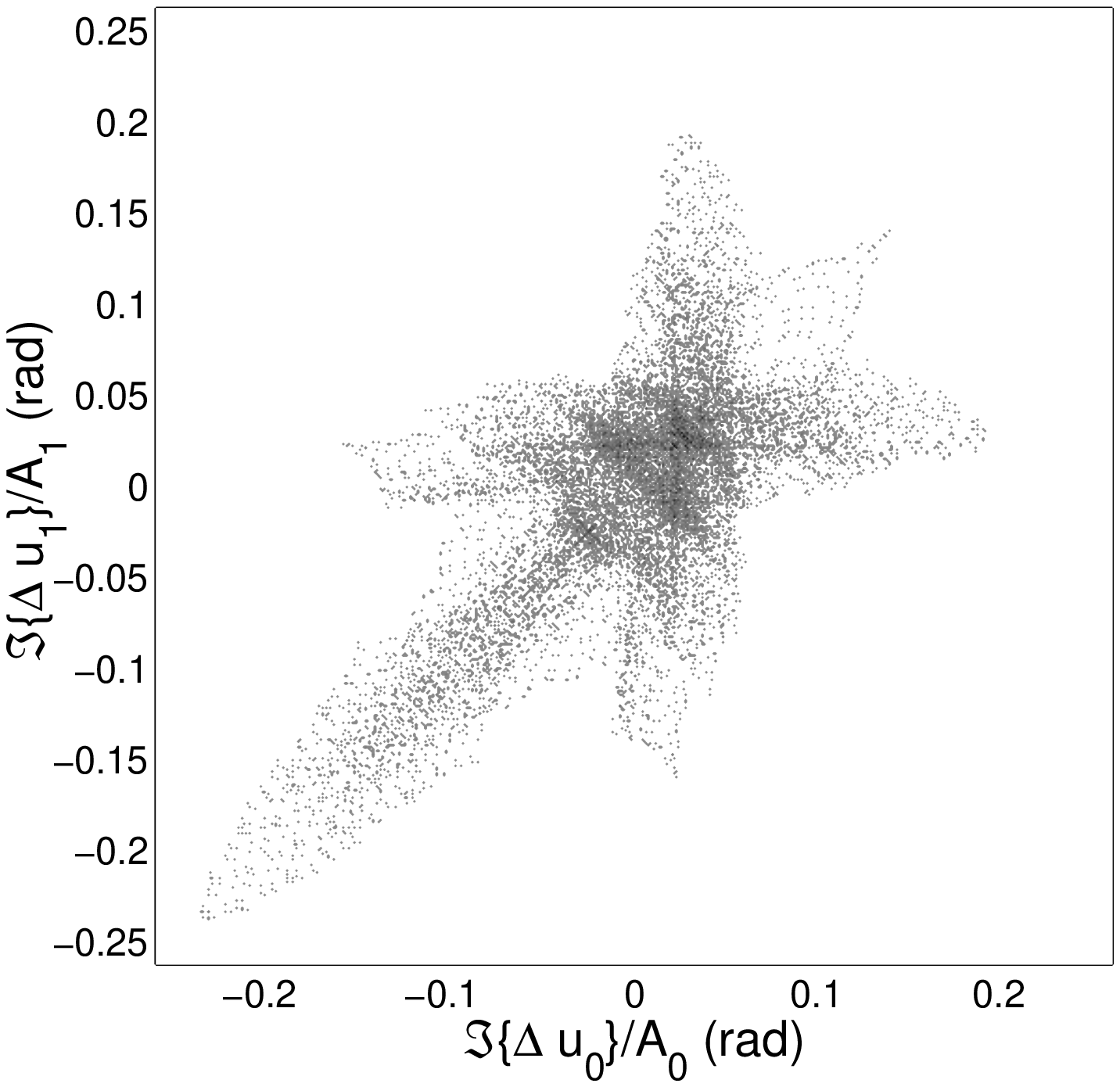} \\
(a) & (b)  
\end{tabular}
\caption{(a) The distribution of the complex electric field of $\Delta u_0/A_0$.
(b) The distribution of $\Im\{\Delta u_0\}/A_0$ versus $\Im\{\Delta u_1\}/A_1$ for two consecutive bit interval.
}
\label{figdist}
\end{center}
\end{figure}

\Fig{figdist}(a) shows the distribution of the normalized complex electric field of $\Delta u_0/A_0$ with the unit of radian.
\Fig{figdist}(b) shows the distribution of the peak phase shift of $\Im\{\Delta u_0\}/A_0$ versus $\Im\{\Delta u_1\}/A_1$ between two consecutive time intervals, where  $\Delta u_0$ and $\Delta u_1$ are the peak-amplitude of ghost pulses and  $\Im\{\cdot\}$ is the imaginary part of a complex number.
The ghost pulses of $\Delta u_0$ and $\Delta u_1$ include all contributions of $-8 < m, n, m + n \leq 8$ for a 16-bit DPSK signal with about 64,000 combinations.
\Fig{figdist}(a) is for ghost pulse at the center bit and \fig{figdist}(b) is for the center two bits. 
If the pulse amplitude of $|A_0|$ is significantly larger than the IFWM ghost pulses, $\Im\{\Delta u_0\}/A_0$ and $\Im\{\Delta u_1\}/A_1$ give approximately the phase shift \cite{wei03fwm}.

\Figs{figdist} are obtained for an $N$-span fiber link with $L =$ 100 km of fiber per span with a normalized launched power of unity mean nonlinear phase shift of $\left<\Phi_\mathrm{NL}\right> = N \gamma L_\mathrm{eff} P_0 = 1$ rad,  where $P_0 = \sqrt{\pi} |A_0|^2 T_0/T$ is the launched power, and $L_\mathrm{eff} \approx 1/\alpha$ is the effective nonlinear length per span.
$N$ identical fiber spans are repeated one after another with 100\% dispersion compensation at the end of each fiber span.
For arbitrary fiber link configuration, $N$ instead of one integration of \eqn{deluifwm} are required.
IFWM ghost pulses add coherently span after span as the worst case. 
The fiber link has an attenuation coefficient of $\alpha = 0.2$ dB/km.
With bit interval of $T = 25$ ps, DPSK signal has a data rate of 40 Gb/s.
The initial pulse width is $T_0 = 5$ ps, for a duty cycle of about $1/3$.
The fiber dispersion is $\beta_2 = -22~$ps$^2$/km, corresponding to $D = 17$ ps/km/nm at the wavelength of $1.55~\mu$m for standard single-mode fiber.

The distribution of \fig{figdist}(a) is very irregular and has significant discrepancy with Gaussian distribution.
Similar to Wei and Liu \cite[Fig. 2]{wei03fwm}, \Fig{figdist}(b) is only symmetrical with respect to $x = y$.
The difference of \fig{figdist}(b) with \cite{wei03fwm} is for a lossy instead of lossless fiber.
With span by span dispersion compensation and for lossy fiber, 50\% precompensation of dispersion increases instead of reduces IFWM.
The phase of $\Im\{\Delta u_0\}/A_0$ is correlated with $\Im\{\Delta u_1\}/A_1$ with a correlation coefficient of about 0.58. 
For $N$ repeated identical fiber spans, \Figs{figdist} are valid for single- and multi-span systems with $\left<\Phi_\mathrm{NL}\right> = 1$ rad.
Note that both $\Delta u_0/A_0$ and $\Delta u_1/A_1$ are zero mean.
Not shown in \figs{figdist}, the real parts of $\Re\{\Delta u_0\}/A_0$ and $\Re\{\Delta u_1\}/A_1$ have a correlation coefficient of about -0.54.

\section{Error Probability for DPSK Signals}

The error probability of DPSK signals with IFWM is difficult to find analytically.
From \figs{figdist}, the distribution of the IFWM induced ghost pulses is not Gaussian distributed.
With the distribution of \fig{figdist}, the error probability of DPSK signal with IFWM can be calculated semi-analytically.

Assumed for simplicity that the transmitted phases at $t = 0$ and $t = T$ are identical and, without loss of generality, the transmitted signals are $E_s(t) = E_s(t - T) = A_0 > 0$.
With optical amplifier noise of $n(t)$, ignored the constant factor of interferometer loss and photodiode responsivity, the photocurrent is \cite{ho0401}
\begin{eqnarray}
i(t) &= & |2A_0 + \Delta u_1 +\Delta u_0 + n(t) + n(t-T)|^2 \nonumber \\
    & & \qquad - |\Delta u_1 - \Delta u_0 + n(t) - n(t - T)|^2.
\label{itifwm}
\end{eqnarray}
 
\noindent A decision error occurs if $i(t) < 0$.

Given $\Delta u_0$ and $\Delta u_1$,  the two terms in \eqn{itifwm} are independent of each other and have a noncentral chi-square distribution \cite[pp. 41-44]{proakis4}.
Each term of \eqn{itifwm} has the same noise variance of $4 \sigma_n^2$ where $E\{|n(t)|^2\} = 2 \sigma_n^2$ with $\sigma_n^2$ as the noise variance per dimension.
The noncentralities of the two terms of \eqn{itifwm} are $|2A_0 + \Delta u_1 +\Delta u_0|^2$ and $|\Delta u_1 - \Delta u_0|^2$, respectively.
From \cite[App. B]{proakis4} \cite{stein64}, the probability of $i(t) < 0$ is equal to

\begin{equation}
p_e(\Delta u_0, \Delta u_1) = Q(a, b) - \frac{1}{2} e^{-(a^2 + b^2)/2} I_0(a b),\label{berDPSKifwm1}
\end{equation}

\noindent where $Q(\cdot, \cdot)$ is the Marcum $Q$ function and
\begin{eqnarray}
a^2 &=& \frac{\rho_s}{2} \left|2 + \frac{\Delta u_1}{A_0} +\frac{\Delta u_0}{A_0} \right|^2, \\ 
b^2 &=& \frac{\rho_s}{2} \left|\frac{\Delta u_1}{A_0} - \frac{\Delta u_0}{A_0} \right|^2,
\end{eqnarray}

\noindent where $\rho_s = A_0^2/2 \sigma_n^2$ is the SNR without taking into account the ghost pulses.
Evaluated simi-analytically using the distribution of \figs{figdist}, the error probability is equal to 

\begin{equation}
p_e = E \left\{ p_e(\Delta u_0, \Delta u_1) \right\},
\label{berDPSKifwm}
\end{equation}

\noindent where $E\{\cdot\}$ denotes expectation.

When the sequence of $A_k$ is changed to $(-1)^k A_k$ with all odd positions changing sign, from \eqn{deluifwm}, $\Delta u_0$ remains the same but $\Delta u_1$ changes sign. 
As $\Delta u_0/A_0$ and $\Delta u_1/A_1$ remain the same for \fig{figdist}(b), the error probability for the case with $A_0 = -A_1$ is the same as that of \eqn{berDPSKifwm}. 

\begin{figure}
\centerline{\includegraphics[width = 0.3 \textwidth]{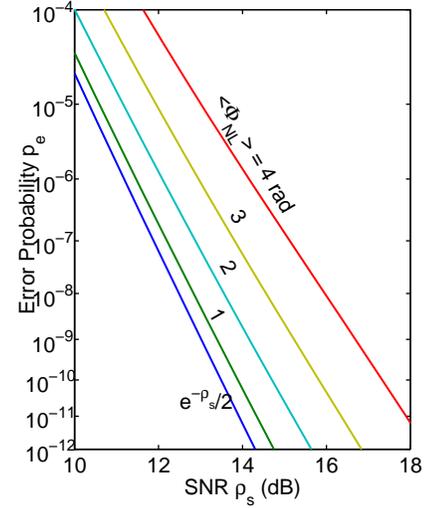}}
\caption{The error probability of DPSK signals with IFWM ghost pulses as a function of SNR $\rho_s$.}
\label{figBerIFWM}
\end{figure}

\Fig{figBerIFWM} shows the error probability as a function of SNR $\rho_s$ for DPSK signal with IFWM induced ghost pulses.
The error probability without IFWM of $\frac{1}{2}e^{-\rho_s}$ \cite[Sec. 5.2.8]{proakis4} is also shown for comparison.
The semi-analytical formula of \eqn{berDPSKifwm} with \eqn{berDPSKifwm1} is used to calculate the error probability based on IFWM ghost pulse distribution of \figs{figdist}.

\begin{figure}
\centerline{\includegraphics[width = 0.38 \textwidth]{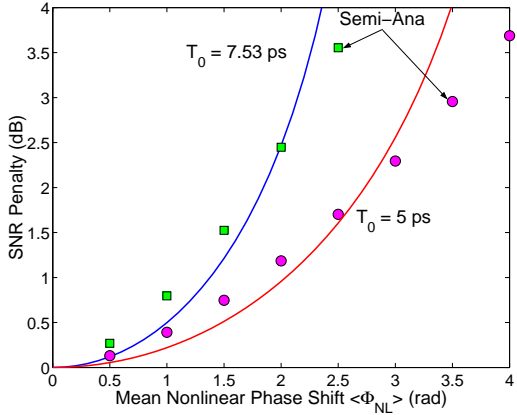}}
\caption{The SNR penalty versus mean nonlinear phase shift of $\left<\Phi_\mathrm{NL}\right>$.
The curves are from empirical formula.}
\label{figPenIFWM}
\end{figure}

\Fig{figPenIFWM} shows the SNR penalty for $p_e = 10^{-9}$ as a function of the mean nonlinear phase shift of $\left<\Phi_\mathrm{NL}\right>$.
In additional to the SNR penalty corresponding to \fig{figBerIFWM}, \Fig{figPenIFWM} also shows the penalty when the initial pulse width is $T_0 = 7.53$ ps for a duty cycle of 1/2.
For 1-dB SNR penalty, the mean nonlinear phase shift must be less than 1.25 and 1.80 rad for initial pulse width of $T_0 = 5$ and $7.53$ ps, respectively.
The SNR penalty is smaller for small initial pulse width of $T_0 = 5$ ps.

If the IFWM induced ghost pulses are assumed as Gaussian distributed, the noise increases to $n(t) + \Delta u_0$ at $t = 0$.
However, the SNR of  $|A_0|^2/E\{|n(t) + \Delta u_0|^2\}$ cannot be used directly to find the error probability due the correlation between the ghost pulses of $\Delta u_0$ and $\Delta u_1$.
The phase variance of $A_0 + n(t)$ is equal to about $1/2\rho_s$ \cite[Sec. 5.2.7]{proakis4}.
The variance of the differential phase is $1/\rho_s$ and should compare with 

\begin{equation}
\sigma^2_{\delta \theta} = E\left\{ \left[\Im\{\Delta u_0/A_0 \}  - \Im\{\Delta u_1/A_1 \} \right]^2 \right\},
\end{equation} 

\noindent due to IFWM ghost pulses.
The SNR penalty can be empirically estimated as $-10\cdot \log_{10} \left( 1 - 20 \sigma^2_{\delta \theta} \right)$, where 20 is the SNR for a DPSK error probability of $10^{-9}$.
This empirical formula finds the SNR penalty based on the variance of the differential phase.
\Fig{figPenIFWM} also shows the SNR penalty from the empirical approximation.
For SNR penalty less than 2 dB, the empirical approximation underestimates the SNR penalty by less than 0.25 dB.
For SNR penalty larger than 2 dB, the approximation overestimates the SNR penalty.

The empirical approximation still requires extensive numerical calculations to find the variance of $\sigma^2_{\delta \theta}$ from many combinations of bit sequence.
The semi-analytical method needs one further step to find the error probability of \eqn{berDPSKifwm1} for each term of $\Delta u_0$ and $\Delta u_1$, and then averaging of \eqn{berDPSKifwm}.

The above analysis and numerical results always used the peak amplitude of the ghost pulses and the signal pulses.
The pulse width of the ghost pulses is ignored for simplicity.
However, the IFWM ghost pulse broadens to $\sqrt{3}$ times the signal pulse width \cite{mecozzi00, mecozzi01a}.
As the power is proportional to the pulse width, the ghosts pulse has $\sqrt{3}$ times larger energy than the signal for the same peak amplitude.
In the worst case, the $x$-axis of \fig{figPenIFWM} must scale by a factor of $3^{1/4} = 1.32$.
However, the scale factor depends on the bandwidth of the optical and electrical filters in the receiver.

If optical match filter precedes the direct-detection DPSK receiver and the electric filter at the receiver has a wide bandwidth that does not distort the signal, the IFWM amplitude is increased by a factor of $\sqrt{3/2} = 1.22$ by the receiver, slightly less than the ratio of $1.32$.
If both the optical and electrical filters have a very wide bandwidth, allowing too much noise to the receiver, the peak amplitude directly transfers to the receiver. 
If the optical filter has a wide bandwidth but the electrical filter is a $0.75/T$ bandwidth Bessel filter, IFWM increases by a factor of 1.19 and 1.33 for $T_0 = 7.53$ and $5$ ps, respectively.
In practical system design, \Fig{figPenIFWM} must be modified to take into account the design of both receiver and transmitter.
Note that the mean nonlinear phase shift of \fig{figPenIFWM} is a simple system parameter to evaluate.

\Fig{figPenIFWM} shows that DPSK signal with IFWM can tolerate a far larger mean nonlinear phase shift of $\left<\Phi_\mathrm{NL}\right>$ than DPSK signal with nonlinear phase noise of $\left<\Phi_\mathrm{NL}\right> < 0.6$ rad for 1-dB penalty \cite{ho0404}.
However, Ho \cite{ho0404} is for return-to-zero (NRZ) signal without pulse distortion and deduces that RZ signal has lower tolerance to nonlinear phase noise.
We are currently developing model for RZ signal with nonlinear phase noise in highly dispersive systems for a fair comparison.

\section{Conclusion}

When the peak amplitude of IFWM induced ghost pulses is evaluated numerically, the error probability of DPSK signals can be found semi-analytically. 
For a SNR penalty less than 1 dB, the mean nonlinear phase shift of the system must be less than 1 to 2 rad depending on the initial pulse width.
An empirical approximation is also used to find the SNR penalty up to 2 dB.


\end{document}